\documentclass[aps,prl,twocolumn,groupedaddress]{revtex4}
\usepackage{epsfig}

\begin{document}

\title{Critical Exponents from Dimensional Variation}

\author{H. Ballhausen}
\email{physics@ballhausen.com}
\affiliation{Institute for Theoretical Physics, Heidelberg University\\Philosophenweg 16, 69120 Heidelberg, Germany}

\begin{abstract}
\noindent
Recent work \cite{Lit1,Lit1a} on exact renormalization group flow equations has pointed out the possibility to study 
critical phenomena in continuous dimension $ D $ of space. In an investigation of the O($N$) model the dimension $ N $
of the fields may be seen as a continuous parameter as well. One may conjecture that a variation of $ D $ or
$ N $ in the vicinity of a second order phase transition yields the same critical exponents as a variation of the 
temperature. A numerical computation confirms this.
\end{abstract}

\maketitle

~

\bigskip
\noindent
\textbf{Dimensional reduction}

\bigskip
\noindent
A well known phenomenon is dimensional reduction. Its most prominent example being the fact
that a quantum field theory in regular space of dimension three in the zero temperature 
limit is equivalent to a quantum field theory in dimension four but in the infinite temperature
limit. This can be derived formally and is easily understood: in thermal field theory the fields
may be expanded into a Fourier series of Matsubara modes due to cyclic boundary conditions. The
circumference of this additional cyclic dimension is inversely proportional to the temperature.

\bigskip
\noindent
Consider then fields in dimension three in the high temperature limit. The additional dimension
is rolled in and all Matsubara modes propagate as a single mode. Lowering the temperature the 
spectrum of Matsubara modes becomes visible. In the limit of zero temperature, finally, the
additional dimension becomes flat and the spectrum of Matsubara momenta becomes continuous. Now
there is effectively a fourth flat dimension. There is no difference to a field theory in space
of dimension four and in the high temperature limit.

~

\bigskip
\noindent
\textbf{Dimensional variation}

\bigskip
\noindent
A conclusion from the picture sketched above may be phrased in the following way:
A system $(D,T)$ in space of dimension $D$ and at temperature $T$ is equivalent to some system 
$(D^\prime<D,T^\prime<T)$. The example above corresponds to $(D=3,T \to 0) \hat{=} (D=4,T \to \infty)$.

\bigskip
\noindent
Consider now second order phase transitions in the O($N$) model. At a critical point $(N_c,D_c,T_c)$
the mass vanishes like $m \sim |T-T_c|^\nu$, it is zero below and nonzero above the critical temperature: 
$m(N_c,D_c,T>T_c)>0$. The above reasoning then leads to the conclusion that there also is a 
dimension $D<D_c$ such that $m(N_c,D<D_c,T_c)$ is nonzero.

\newpage
~

\bigskip
\noindent
~

\bigskip
\noindent
Simulations show that this holds for arbitrarily small deviations from the critical point $(N_c,D_c,T_c)$.
That is why we conjecture that there exists in a small vicinity of the critical point
a continuous function $f(\epsilon)$ such that $m(N_c,D_c,T_c+\epsilon)=m(N_c,D_c-f(\epsilon),T_c)$:

\bigskip
\noindent
\begin{center}
\epsfig{file=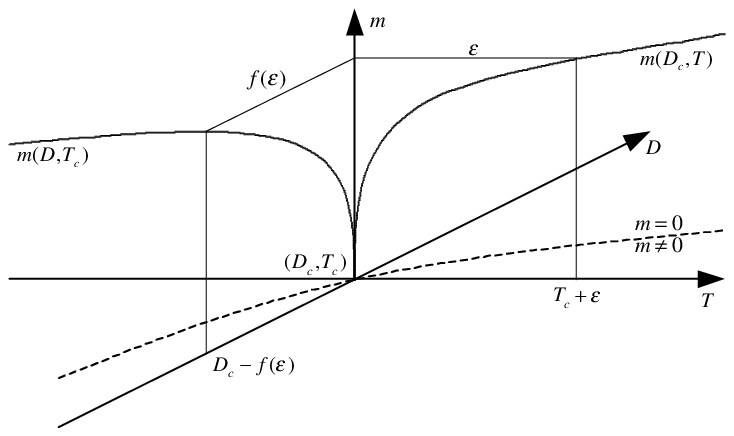,width=8.5cm,bbllx=0,bblly=245,bburx=240,bbury=370}
\end{center}

\bigskip
\noindent
For small $\epsilon$ $f(\epsilon)$ may be expanded into a power series: $f(\epsilon)=f_0+f_1 \epsilon+...$,
hence $m(N_c,D_c,T_c+\epsilon)=m(N_c,D_c-f_0-f_1 \epsilon-...,T_c)$. At the critical point the mass is zero, 
just above it is nonzero, thus $f_0=0$. Then holds close to the critical point 
$m(N_c,D_c,T_c+\epsilon)=m(N_c,D_c-f_1 \epsilon,T_c)$. Taking the limit $m \to 0$, $\epsilon \to 0$
it follows from $\epsilon \sim f_1 \epsilon$ that $T-T_c \sim D_c-D$ and hence \\
$m \sim |T-T_c|^\nu \sim |D-D_c|^\nu$.

\bigskip
\noindent
Observation suggests that the same discussion holds for a variation of the dimension of the fields $N$
around $N_c$. We thus expect to find the same critical exponent $\nu$ either by a variation of the temperature
or by dimensional variation of $D$ or $N$. This has recently also found an explanation in terms of analyticity
properties of the $\beta$-function \cite{Lit1} and can be confirmed using exact renormalization group flow equations:

\newpage
\noindent
\textbf{Exact Renormalization Group Flow Equations}

\bigskip
\noindent
The effective average action ( see e.g. \cite{Lit2} ) $\Gamma_k$ interpolates between the classical microscopic action $ S $ at the ultraviolet
scale $ k=\Lambda $ and the effective action $ \Gamma $ at $ k=0 $. It contains all fluctuations to arbitrarily high
loop order with momenta larger than $ k $. Its flow is described by a non-perturbative exact renormalization
group flow equation:

\bigskip $ \partial_k \Gamma_k = \frac{1}{2} \textrm{Tr} \left( \partial_k R_k \cdot \left[ \Gamma^{(2)}_k + R_k \right]^{-1} \right) $

\bigskip
\noindent
In a derivative expansion

\bigskip $ \Gamma_k = \int \frac{d^Dq}{(2\pi)^D} (U_k(\rho)+q^2\rho Z_k+q^2\rho^2Y_k) $
 
\bigskip
\noindent
 where $\rho=\frac{1}{2}\varphi^2$ the flow of the potential reads:

\bigskip $ \partial_k U_k = \frac{1}{2} \int \frac{d^Dq}{(2\pi)^D} \cdot \partial_k R_k(q^2) \cdot \\ ~~~~~~~~~ \left( \frac{1}{U_k^\prime+2\rho U_k^{\prime\prime}+q^2Z_k+q^2\rho Y_k+R_k} + \frac{N-1}{U_k^\prime+q^2Z_k+R_k} \right) $

\bigskip
\noindent
The radial renormalization $\tilde{Z}_k=Z_k+\rho Y_k$ and the Goldstone renormalization 
$Z_k$ may be combined in such a way into $\bar{Z}_k$ that the anomalous dimension reads 
$ \bar{\eta} = \partial_t \textrm{Ln} \bar{Z_k} = (\tilde{\eta}+(n-1)\eta)/n$,
where $ t = \textrm{Ln}(k/\Lambda) $ and $n=\textrm{Min}(1,N)$. 

\bigskip
\noindent
Choosing $ R_k(q^2)=\bar{Z}_k(k^2-q^2)\Theta(k^2-q^2) $ \cite{Lit3} as the infrared regulator, and 
switching to dimensionless and renormalized quantities, the remaining two flow equations read:

\bigskip $ \partial_t u = - D u + (D-2+\bar{\eta})\rho u^\prime + 2 v_D \cdot \\ ~~~~~~~~~~~~~~~~~~~~~ \left( \frac{2-\bar{\eta}}{D}+\frac{\bar{\eta}}{D+2} \right) \left( \frac{1}{1+u^\prime+2\rho u^{\prime\prime}} + \frac{N-1}{1+u^\prime} \right) $

\medskip $ \bar{\eta} = \frac{8 v_D}{D} \frac{\kappa u^{\prime\prime 2}}{n} \cdot \\ ~~~~~~~~~~~~~~~~~~~~~ \left( \frac{(3+2\kappa u^{\prime\prime\prime}/u^{\prime\prime})^2}{(1+2\kappa u^{\prime\prime})^4} + \frac{2(n-1)}{(1+2\kappa u^{\prime\prime})^2}+ \frac{N-1}{1} \right) $

\bigskip
\noindent
where $ v_D^{-1} = 2^{D+1} \pi^{D/2} \Gamma(D/2) $ and $ \kappa $ is the running minimum of $ u $. 
Starting with a quartic potential $ u = \frac{1}{2}(\rho - \kappa_\Lambda)^2 $ 
the flow equations are discretized on a grid and numerically solved for different values of 
$ \kappa_\Lambda $. During the evolution towards $ k \to 0 $, the running minimum of the 
potential (~$u^\prime(\kappa)=0$~) may either end up in the massless spontaneously broken phase 
(~$m=0$, $\kappa \to \infty $~) or in the massive symmetric phase (~$ m^2 \sim k^2 u^\prime(0)$, $\kappa = 0 $~). 

\bigskip
\noindent
The critical $ \kappa_c $ leads to a second order phase transition characterized by fix points of all couplings and vanishing masses.
Fine tuning $ \kappa_\Lambda $ around $ \kappa_c $ then yields the critical exponent $ \nu $ through the scaling law $ m \sim | T-T_c |^\nu $
and the established proportionality $ | \kappa_\Lambda - \kappa_c | \sim | T - T_c | $.

\bigskip
\noindent
The critical exponent $ \eta $ is given by the fix point of the anomalous dimension $ \bar{\eta} $. It therefore only depends on the
critical trajectory and is thus the same for dimensional variation. All other exponents follow from scaling laws. So it suffices to
show that the exponent $ \nu $ is the same for variation of temperature and dimensional variation:

\bigskip
\noindent
~

\bigskip
\textbf{Numerical Results}

\bigskip
\noindent
The following table gives the critical exponent
$\nu$ for different values of $N_c$ and $D_c$.
It lists the numerical values for a variation of
the temperature and for dimensional variation and the
corresponding values given by literature ( see e.g. \cite{Lit4} ).

\bigskip
\begin{tabular}{|cc|cccc|}
\hline
~~$N_c$~&~$D_c$~~&~$\delta T$~&~$\delta D$~&~$\delta N$~&~Lit.~\\
\hline
~~~0~~&~3~~~&~~~0.5895~~&~~0.5900~~&~~0.5891~~&~~0.588~~~\\
~~~0~~&~4~~~&~~~0.5003~~&~~0.5014~~&~~0.5003~~&~~0.500~~~\\
~~~1~~&~2~~~&~~~1.0535~~&~~1.0535~~&~~1.0535~~&~~1.000~~~\\
~~~1~~&~3~~~&~~~0.6270~~&~~0.6273~~&~~0.6270~~&~~0.631~~~\\
~~~1~~&~4~~~&~~~0.5004~~&~~0.5002~~&~~0.5004~~&~~0.500~~~\\
~~~2~~&~3~~~&~~~0.6665~~&~~0.6665~~&~~0.6665~~&~~0.671~~~\\
~~~2~~&~4~~~&~~~0.5006~~&~~0.5006~~&~~0.5005~~&~~0.500~~~\\
~~~3~~&~3~~~&~~~0.7026~~&~~0.7025~~&~~0.7026~~&~~0.710~~~\\
~~~3~~&~4~~~&~~~0.5007~~&~~0.5007~~&~~0.5008~~&~~0.500~~~\\
~~~4~~&~3~~~&~~~0.7341~~&~~0.7342~~&~~0.7340~~&~~0.739~~~\\
~~~4~~&~4~~~&~~~0.5009~~&~~0.5009~~&~~0.5009~~&~~0.500~~~\\
\hline
\end{tabular}

\bigskip
\noindent
All values are in good agreement with literature.
More importantly, the agreement between the values given by
the three methods is excellent.
One may conclude that dimensional variation yields the very same
critical exponents as the usual variation of temperature does.

\end{document}